\documentclass[12pt,preprint]{aastex}
\usepackage{t1enc}
\usepackage{graphicx}
\usepackage{epstopdf}
\usepackage{natbib}
\usepackage{lscape}
\newcommand{\gcvn}{G5.89-0.39}
\newcommand{\gcns}{G5.97-1.17}
\newcommand{\gunsu}{G19.61-0.23}
\newcommand{\gdzzv}{G20.08-0.14}
\newcommand{\gdvdn}{G28.29-0.36}
\newcommand{\gtcdz}{G35.20-1.74}
\newcommand{\gszvv}{G60.88+0.13}
\newcommand{\gsuqv}{G61.48+0.09}
\newcommand{\gssuv}{G76.18+0.13}
\newcommand{\gsstv}{G76.38-0.62}
\newcommand{\gsvqq}{G78.44+2.66}
\newcommand{\gsntz}{G79.30+0.28}

\newcommand{\msun}{M$_{\odot}$}

\newcommand{\cmt}{cm$^{-3}$}

\newcommand{\jpb}   {$\rm Jy~beam^{-1}$}

\newcommand{\gap}%
{\raisebox{-0.5ex}{$\stackrel{\scriptstyle >}{\scriptstyle \sim}$}}

\newcommand{\kms} {km~s$^{-1}$}

\begin{document}

\shorttitle{Compact sources in UCHII regions}
\shortauthors{Masqu\'e et al.}

\title{Searching for compact radio sources associated to UCHII regions }

\author{
Josep M. Masqu\'e\altaffilmark{1},
Luis F. Rodr\'iguez\altaffilmark{2,3},
Miguel A. Trinidad\altaffilmark{1},
Stan Kurtz\altaffilmark{2},
Sergio A. Dzib\altaffilmark{4},
Carlos A. Rodr\'iguez-Rico\altaffilmark{1}, \&
Laurent Loinard\altaffilmark{2}
} 

\altaffiltext{1}{Departamento de Astronom\'ia, Universidad de Guanajuato, Apdo. Postal 144, 36000 Guanajuato, M\'exico}
\altaffiltext{2}{Instituto de Radioastronom\'ia y Astrof\'isica, Universidad Nacional Aut\'onoma de M\'exico, Morelia 58089, M\'exico}
\altaffiltext{3}{Astronomy Departament, Faculty of Science, King Abdulaziz University, P.O. Box 80203, Jeddah 21589, Saudi Arabia}
\altaffiltext{4}{Max Planck Institut f\"ur Radioastronomie, Auf dem H\"ugel 69, D-53121 Bonn, Germany}
\begin{abstract}

Ultra-Compact (UC)HII regions represent a very early stage of massive star formation whose structure and evolution are not yet fully understood. Interferometric observations in recent years show that some UCHII regions have associated compact sources of uncertain nature. Based on this, we carried out VLA 1.3 cm observations in the A configuration of selected UCHII regions in order to report additional cases of compact sources embedded in UCHII regions. From the observations, we find 13 compact sources associated to 9 UCHII regions. Although we cannot establish an unambiguous nature for the newly detected sources, we assess some of their observational properties. 
According to the results, we can distinguish between two types of compact sources.  One type corresponds to sources that probably are deeply embedded in the dense ionized gas of the UCHII region. These sources are being photo-evaporated by the exciting star of the region and will last for 10$^4-10^5$ yr. They may play a crucial role in the evolution of the UCHII region as the photo-evaporated material could replenish the expanding plasma and might provide a solution to the so-called \emph{lifetime problem} for these regions. The second type of compact sources is not associated with the densest ionized gas of the region. A few of these sources appear resolved and may be photo-evaporating objects such as those of the first type but with significantly lower mass depletion rates. The rest of sources of this second type appear unresolved and their properties are varied. We speculate on the similarity between the sources of the second type and those of the Orion population of radio sources.


\end{abstract}

\section{Introduction}

In their early stages, massive stars produce compact photo-ionized regions with small sizes ($\leq0.1$~pc) and large electron densities and emission measures (\gap~10$^4$~cm$^{-3}$, and \gap~10$^7$~pc~cm$^{-6}$, respectively) known as ultra-compact (UC) HII regions (see \citealt{churchwell1990}  and \citealt{kurtz2002}). 
These regions are believed to represent a very early stage of massive star formation as they are deeply embedded in molecular clouds and still retain high density (ionized) gas. The high pressure contrast between the ionized gas and the neutral surrounding medium leads to the common assumption that UCHII regions are freely expanding at the sound speed of ionized gas \citep[$\sim 10$ \kms,][]{spitzer1978}. Thus, their material should disperse very fast becoming significantly more extended in $\sim10^4$ yr. However, \citet{wood1989} carried out a survey of selected sources in the Galactic plane and detected a much large number of UCHII regions than expected for their short lifetime.


Various models have been proposed to explain this paradox. Some of them describe mechanisms to confine the ionized region  \citep[e.g.][]{vanburen1990,xie1996} while others are based on the continuous replenishment of the ionized gas \citep[e.g.][]{tenoriotagle1979,hollenback1994}. The prime mechanism for the confining models is the presence of an ionized accreting flow that quench the expansion of the ionized gas \citep[e.g.,][]{galvanmadrid2011}. Observationally, a variable ionized accretion flow can produce flux density variations, as recently reported in some UCHII regions \citep{francohernandez2004, galvanmadrid2008, depree2015}. However, other authors attribute these flux variations to changes in the source of ionizing radiation \citep[e.g.,][]{francohernandez2004,klassen2012a,klassen2012b}. On the other hand, the \emph{replenishment} models suggest substructure within UCHII regions such as density gradients or photo-evaporating disks \citep[e.g.][]{yorke1996}. There is ample evidence that the homogeneous Str\"omgren sphere is an oversimplification not suitable to fully explain the evolution of UCHII regions. For this purpose, investigations of the small-scale structure of UCHII regions are of paramount importance, as they may reveal the sources of gas replenishment required by some models to explain the lifetime of UCHII regions.



 In recent years, several compact sources possibly associated with UCHII regions have been reported \citep{kawamura1998, dzib2013a,carral2002, rodriguez2014}. A naive interpretation is that these sources are the so-called Hyper-Compact (HC)HII regions surrounding even younger neighboring stars.\footnote{HCHII regions constitute the smallest HII regions observed, even though only a handful of this class of objects are known \citep[e.g.][]{sanchez-monje2011b}. They have extremely small sizes ($\lesssim0.03$ pc), high emission measures ($>10^{10}$ pc cm$^{-6}$) and electron densities \citep[$>10^6$ \cmt, ][]{kurtz2005}.} However, very few of these sources have properties compatible with those of HCHII regions. The studies mentioned above show that there is no unique explanation for the nature of these compact sources as they  do not constitute a homogenous class of objects. For example, W3(OH) harbors a compact source that exhibits time variability and a positive spectral index consistent with the presence of a fossil photo-evaporating disk \citep{dzib2013a}. On the other hand, a compact source found in NGC 6334A has a negative spectral index characteristic of optically thin synchrotron emission and it is also time variable \citep{rodriguez2014}.  The synchrotron emission implies the presence of mildly relativistic electrons. A negative spectral index also can indicate gyrosynchrotron emission. In this case the electrons are accelerated in magnetic reconnection events near the surfaces of low-mass young stars. Alternatively, the electron acceleration can also occur in shocks of stellar winds between O companions forming a massive binary system ionizing the HII region  \citep[e.g. Cyg OB2:][]{blomme2013}.  Additional examples where the flux of the source appears constant are found in G5.97-1.17 \citep{masque2014}. An interesting case is  the Monoceros UC HII region where two compact sources have been observed close to its center; one is related to a slightly resolved thermal source \citep{gomez2000} and the other to a magnetically active star \citep{dzib2016}. This indicates that even in the same region compact sources with different nature can co-exist.

A more extensively studied case is the Orion Nebula Cluster (ONC), which shows a large number and variety of compact radio sources reported in the last three decades \citep{garay1987, churchwell1987,zapata2004, kounkel2014, forbrich2016}. The large number of compact sources associated with the Orion Nebula might be partly due to its proximity and to the richness of the cluster. This enables the detection of weak sources (that would remain undetectable in more distant objects), and the resolution of these sources into individual compact components.
In addition, most of the objects present IR, optical or X-ray counterparts depending on their intrinsic nature or extinction. All these properties are indicative of the varied class of objects co-existing in the ONC, making evident       
the striking stellar multiplicity inherent in the vicinity of massive stars.


Motivated by the possible high occurrence of UCHII regions with associated compact sources, we carried out high resolution VLA observations of selected UCHII regions to search for additional cases of compact sources hidden in the bright free-free emission of the photo-ionized region. By inspecting these compact sources in very young environments such as UCHII regions, we are possibly studying the youngest objects of ONC-like radio source populations. The article is structured as follows: in Sect. 2 we describe the observations. In Sect. 3 we present the resulting maps and the newly discovered compact sources found in UCHII regions. In Sect. 4 we analyze some selected sources. Finally, in Sect. 5 we discuss the general results of the analyzed sources and draw conclusions.


\section{Observations}

The 1.3 cm Jansky Very Large Array (JVLA) observations were conducted in two observing blocks on 2014 May 16th and 28th with the A configuration\footnote{The observations presented here were part of NRAO program VLA 14A-481. The National Radio Astronomy Observatory is a facility of the National Science Foundation operated under cooperative agreement by Associated Universities, Inc.}. We observed, in two Observing Blocks, 12 out of 17 proposed UCHII regions of the \citet{wood1989} and \citet{kurtz1994} surveys selected because they showed evidence of compact structures in archival data. The 5 remaining sources correspond to a third Observing Block, which was not observed due to time constraints. We observed half of the source set in each observing block. In the source selection, we also applied the following two criteria: (i) sources must be located at $< 4.5$ kpc and (ii) sources with irregular morphologies are discarded. The studied regions and observational parameters are listed in Table \ref{regions_observed}. We observed the frequency range 18-26 GHz with a spectral setup optimized for continuum observations. We used the 3 bit samplers with full polarization mode and set 2 s as the integration time. The flux calibrator was 3C 286. Pointing corrections were applied for telescope slews larger than 10 degrees on the sky. To obtain a homogeneous $uv-$coverage, the fields were observed with 1 to 1.5 minutes of time on source preceded and followed by 30 seconds observing the gain calibrator.  This sequence was applied to each field and was repeated until an observing block of two hours was completed. This arrangement gives on-source observing times between 4 and 6 minutes.

Data were calibrated with the Common Astronomical Software Applications (CASA) package through the pipeline provided by the National Radio Astronomy Observatory (NRAO). A first set of maps was made with the CLEAN task of CASA with natural weighting and fitting the frequency dependence across the bandwidth with two terms of the Taylor polynomial during the deconvolution (\emph{nterms} parameter set to 2). As we aimed to detect compact components associated with resolved UCHII regions, we obtained a second set of maps of selected sources using uniform weight
and suppressing the bright extended free-free emission by removing the short spacings of the visibilities in most of the regions (between 200 and 800 k$\lambda$ depending on the region). The parameters of both sets of maps are given in Table \ref{maps}. The choice of the $uv$ range was based on minimizing the extended emission without affecting significantly the image quality. The corresponding angular scale for which larger structures are suppressed is given in column 4 of Table \ref{maps}.

\section{Observational Results}
  
\begin{figure}[thbp]
\vspace{-3.6cm}
\resizebox{1.1\textwidth}{!}{\includegraphics[angle=0]{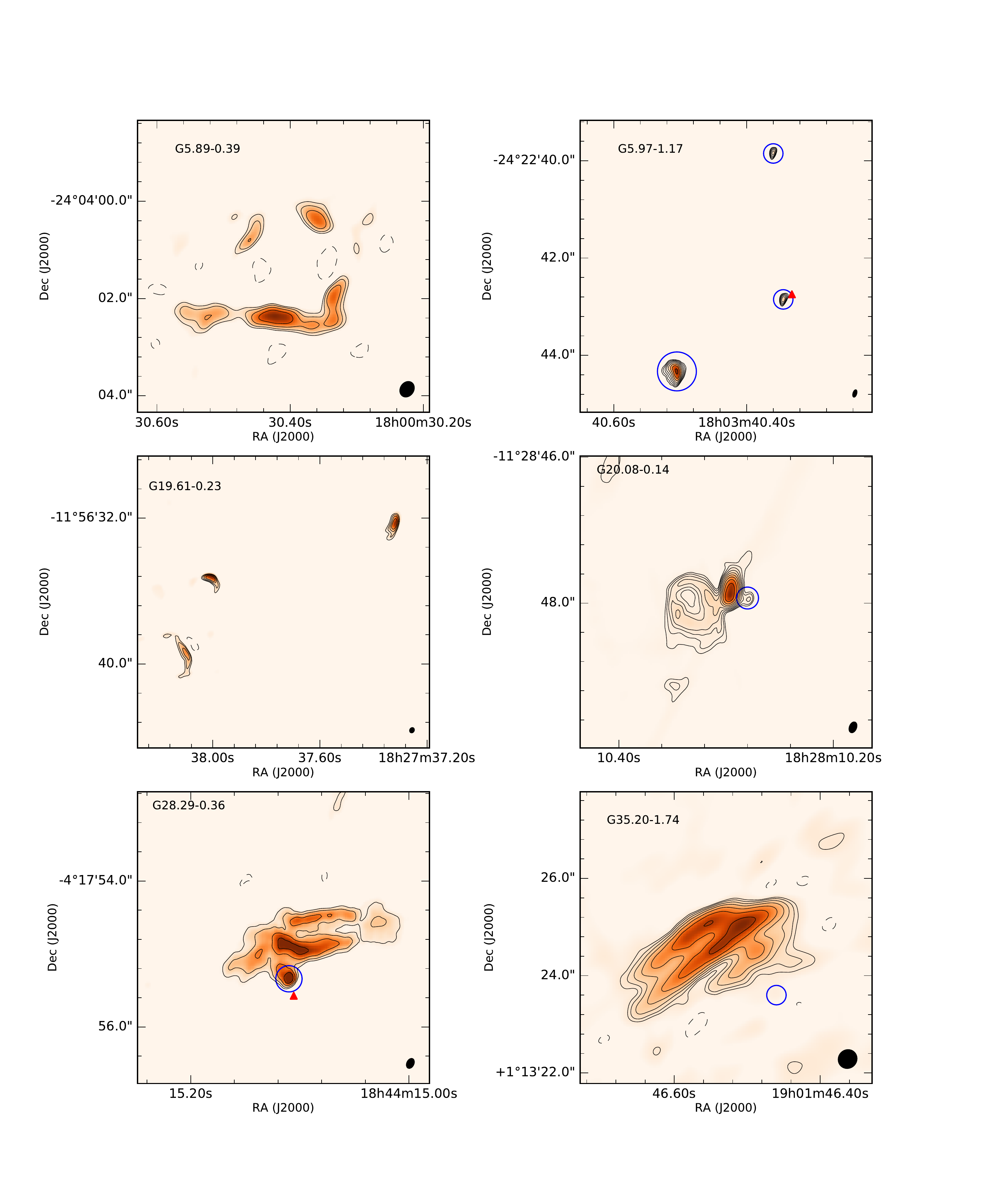}}
\vspace{-1.9cm} 
\caption{
Continuum 1.3 cm emission maps with natural weighting of the observed regions in the Observing Block executed on 2014  May 28th (color scale and contours). For all panels, levels are $C$ times $-2^{0}, 2^{0}, 2^{1/2}, 2^{1}, 2^{3/2},2^{2}, 2^{5/2}, 2^{3}, 2^{7/2}$ and $2^{4}$, where $C$ is given by 2.5 times the $rms$ noise shown in column 3 of Table \ref{maps}, except for \gcvn, \gunsu\ and \gtcdz\ that were constructed with a Gaussian taper of 0.25, 0.2 and 0.35 arcsec, giving $rms$ noise levels of 8.5, 1.5 and 2 m\jpb, respectively.     
The field of view depends on the panel. The synthesized beam is shown in the bottom right corner. The blue circles indicate the compact sources reported in Table \ref{sources} found after removing the short spacings of the visibility dataset. The red triangles indicate the 2MASS (PSC) positions.  \label{general}}
 \end{figure}

\begin{figure}[thbp]
 \vspace{-1.5cm} 
\resizebox{1.1\textwidth}{!}{\includegraphics[angle=0]{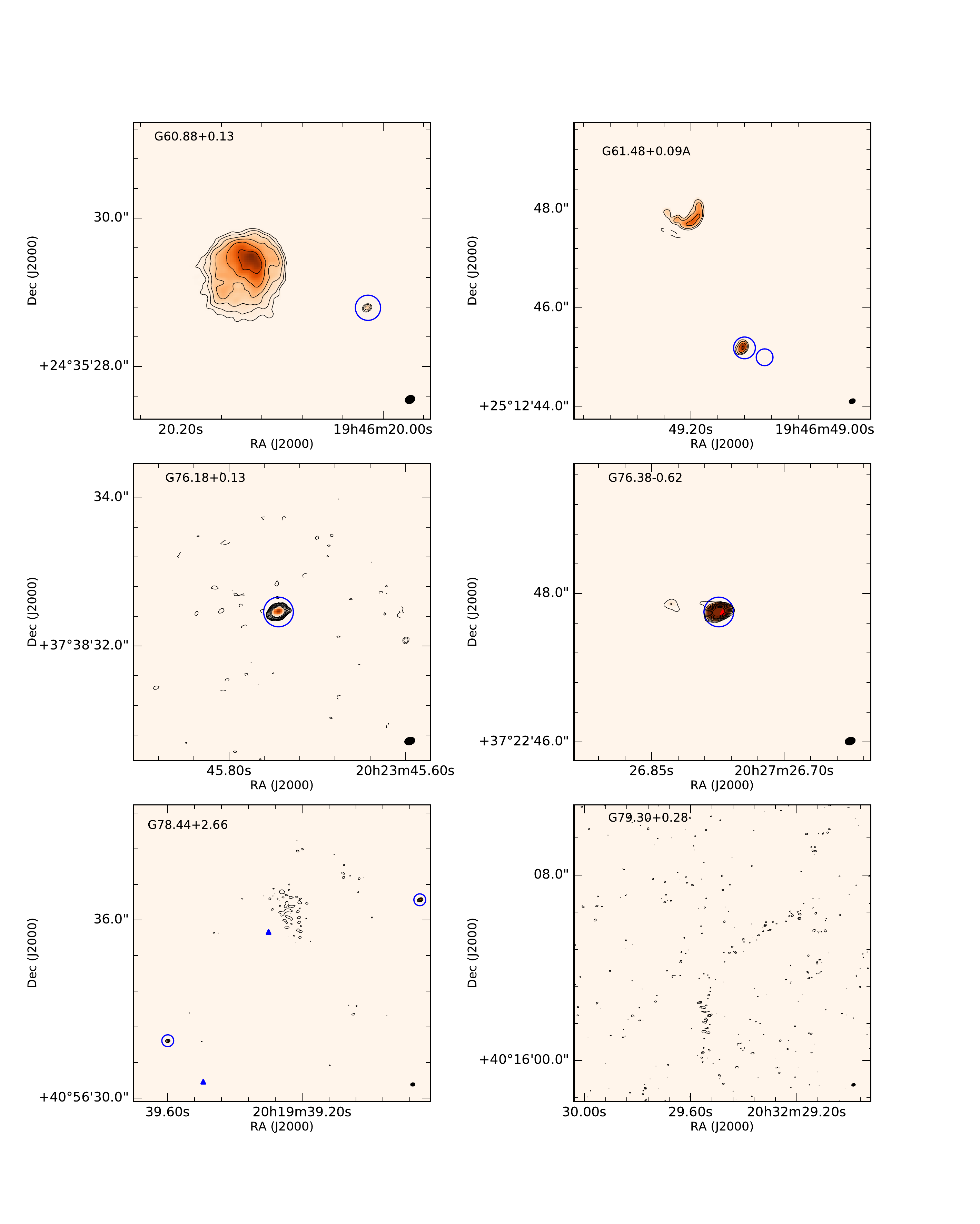}}
\vspace{-2.0cm} 
 \caption{Continuum 1.3 cm emission maps with natural weighting of the observed regions in the Observing Block executed on 2014 May 16th. Contours and symbols are the same as in Figure \ref{general}. In the \gsvqq\ panel the blue triangles show sources number 2 and 1 of \citet{tej2007}, respectively. In this panel, source 2 of \citet{tej2007} is also coincident with the 2MASS source. \label{general_bis}}
\end{figure}

The 1.3 cm continuum emission of the observed regions is shown in Figures \ref{general} and \ref{general_bis}. Due to the lack of short baselines of the A configuration of the VLA, the extended emission of these regions is poorly imaged. This limitation is not important for our study which focuses on the most compact components of the regions. By comparing our 1.3 cm maps with those obtained at 2 cm with the VLA in the B configuration of \citet{wood1989} and \citet{kurtz1994} we assess that, in many regions, only the brightest part of  the emission is recovered. The \gsvqq\ and \gsntz\ regions show the most extreme cases of filtering where practically all the emission disappears.  On the other hand, \gszvv\ and \gsuqv\ present angular sizes $\lesssim\ 1$ arscec and can be fully mapped, while \gssuv\ and \gsstv\ lack any extended structure.

\begin{figure}[thbp]
\vspace{-3.6cm} 
\resizebox{1.0\textwidth}{!}{\includegraphics[angle=0]{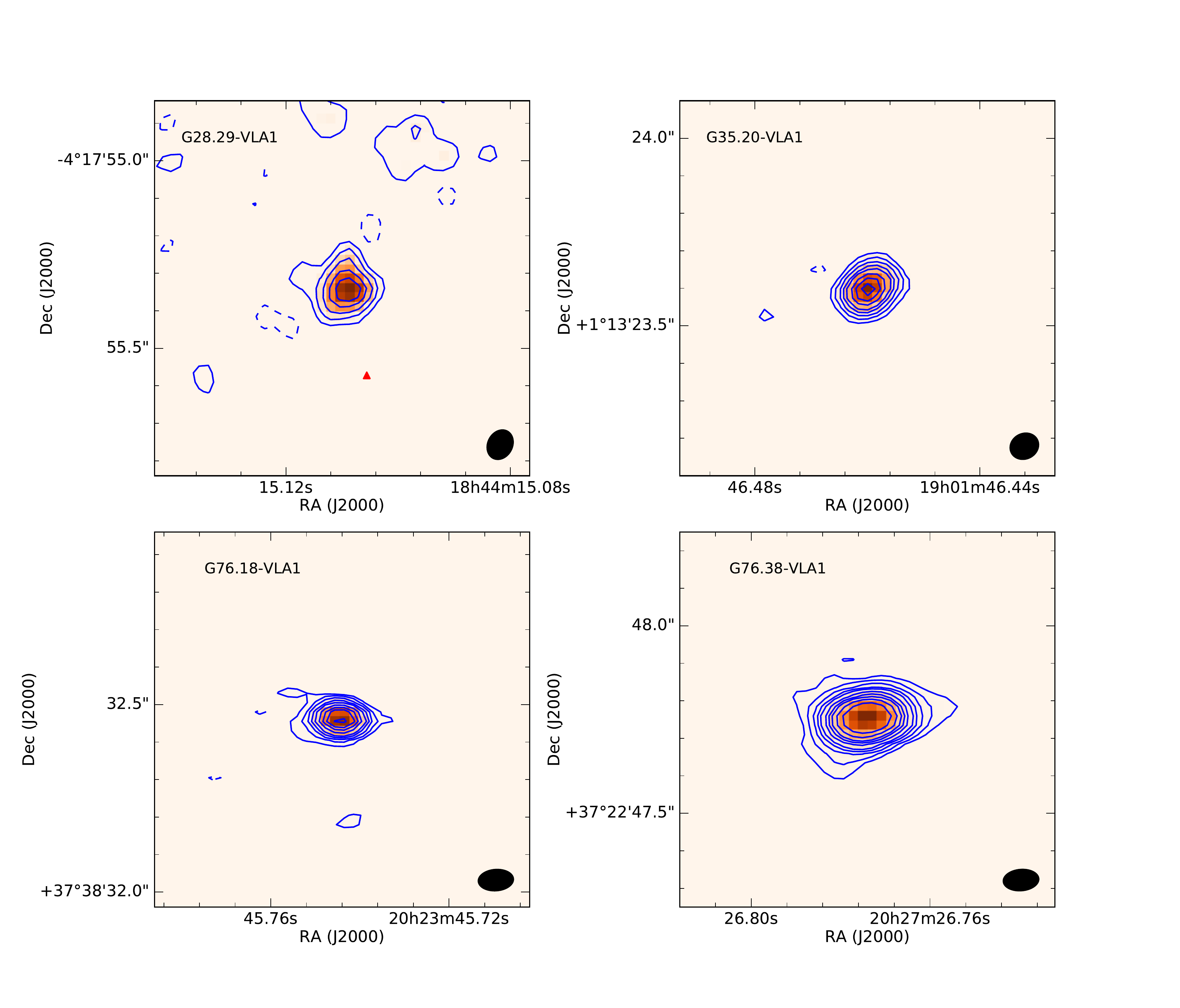}}
\caption{
Close-up continuum maps (color scale and contours) of compact sources of Type I (i.e. embedded in the UCHII region) constructed with uniform weighting and removing the short spacings of the visibility dataset for selected regions. The corresponding largest angular structure mapped is shown in Column 4 of Table \ref{maps}. For all panels, levels are  -3, 3, 6, 10, 15, 20, 30, 40, 50, 70, and 100 times the $rms$ noise shown in column 6 of Table \ref{maps}. The synthesized beam is shown in the bottom right corner. The 2MASS (PSC) position is marked as a red triangle in the G28.29-VLA1 panel. \label{compact_a}}
\end{figure}

We report the presence of compact sources associated with UCHII regions, in most of the cases for the first time, whose positions are indicated in Figs.  \ref{general} and \ref{general_bis}. Since our observations filter out most of the extended emission, the compact sources appear visible at first glance. In some maps, however, the extended emission produces artifacts that reduce the dynamic range and limit our ability to detect compact sources. To  supress the extended emission of the ionized gas around the compact source we removed the short baselines of the visibility dataset for selected regions (see Sect. 2). A close-up view of the detected compact sources from this new set of maps can be seen in Figures \ref{compact_a} and \ref{compact_b}.

Given the small field of view presented in these figures, the probability that a background compact source is seen in the same direction as the UCHII regions is extremely small: following expression A1 of \citet{anglada1998} for a 10\arcsec\  field of view (the largest field of view among all the maps shown in Figs. \ref{general} and \ref{general_bis}) and adopting 3 times the $rms$ noise of 50 $\mu$Jy (typical $rms$ noise as seen in column 6 of Table \ref{maps}) as the detection threshold, we obtain $1 \times 10^{-4}$ expected background sources to be detected at 1.3 cm. The position of the compact source does not necessarily coincide with the brightest part or with the centroid of the UCHII region, as can be inferred from a comparison with larger scale maps found in the literature. To ensure that these compact sources are not merely the peaks of more extended emission, we compared their fluxes and sizes in maps obtained with several $uv$ ranges. Among all the maps of a set corresponding to a specific region, the properties of the compact source barely changed, indicating that they are distinct compact structures. There is, however, the case of \gdvdn, where the compact source appears coincident with the brightest UCHII peak and marginally resolved, making its independent nature from the extended free-free emission questionable.

The observational parameters of the compact sources derived from Gaussian fits are shown in Table \ref{sources}. Their measured flux densities are typically a few mJy and some of them appear marginally resolved with a typical deconvolved size of a few hundreds of AU.  However, due to the limited $uv$ coverage of our observations, some of the deconvolved sizes must be taken with caution.  We attempted to derive spectral indices using the 8 GHz bandwidth but, except for G76.38$-$0.62 and G76.18$+$0.13,
the signal to noise ratio of the sources was insufficient to get trustworthy results.

From the physical size and flux density, we can derive, under certain assumptions, several physical parameters listed in Table \ref{physical_parameters}. The assumptions and the formulas used are explained in the footnotes of the Table. We considered only the physics of the most likely scenario to explain the nature of the compact sources. 
Regions \gssuv\ and \gsstv\ were discarded from the table because they have brightness temperatures of  $25,000 \pm 11,000$ and $15,000 \pm 300$ K, possibly indicative of a hot, optically-thick HCHII region. However, the spectral indices of these sources (see below) do rule out an optically-thick free-free nature. We will dedicate a brief discussion to these sources below.

\section{Analysis of selected sources}

By inspecting Figs. \ref{general} and \ref{general_bis}, we find that the compact sources follow two main trends according to their location with respect to the whole UCHII region. On one hand, there are sources coincident with or close to the centimeter emission peak of the UCHII region (hereafter, sources of Type I). Despite the lack of extended emission of the \gssuv\ and \gsstv\ regions, we include their associated compact sources in this group as they are coincident with the peak emission of the \citet{kurtz1994} maps. On the other hand, there are sources that appear scattered around the UCHII region (hereafter, sources of Type II).

\subsection{Sources of Type I}

 Sources of Type I (i.e. those that are located within the boundaries of the radio emission from the UCHII region) are related to the exciting star of the region given their association with the dense ionized gas. Below we give some individual remarks on the sources of Type I and provide a tentative interpretation for the nature of the compact sources belonging to this group.      

\begin{itemize}

\item[-] \emph{\gdvdn.}          This region is embedded in a dusty massive cloud as indicated by millimeter and submillimeter  observations  \citep{hill2005, rathborne2006,walsh2003}.  
The cloud is coincident with the source IRAS 18416$-$0420 and has extended mid-IR emission that implies the presence of warm dust \citep{walsh2003,rathborne2006}. Moreover, the association of bright radio emission suggests ongoing massive star formation over this region \citep{helfand2006}. The centimeter continuum maps of \citet{kurtz1994} resolve \gdvdn\ into two peaks, aligned north-south, surrounded by diffuse emission. As seen in Figure \ref{g28_29}, within the astrometric uncertainties of the VLA, the  G28.29-VLA1 source is coincident with the northern and brightest peak of the \citet{kurtz1994} map. 
 
 The 2MASS point source catalogue (PSC) reports two IR sources possibly associated with the northern and southern peaks (see Figure \ref{g28_29}). This coincidence suggests that the \gdvdn\  UCHII region is composed of at least two subregions, each one being photo-ionized by its own star, detected at IR wavelengths. A close inspection of the corresponding panel of Figure \ref{compact_a} reveals that the IR source falls $0\rlap{.}''25$ south of G28.29-VLA1. Given the 2MASS PSC astrometric accuracy of $\leq0\rlap{.}''1$, similar to the accuracy of the VLA data, the separation between G28.29-VLA1 and the IR source corresponds to more than two times the $\sigma_{\mathrm{rms}}$ error in the determination of positions with our observations. In addition, G28.29-VLA1 is the largest source of Type I ($\sim300$ AU, see Table \ref{physical_parameters}) and shows some evidence of being extended.

\item[-] \emph{\gtcdz.} The \gtcdz\ region belongs to the W 48 complex found at the edge of a molecular cloud \citep{zeilik1978}. The region presents strong maser activity \citep{genzel1977,evans1979,caswell1995} that led \citet{zhang2009bis}, from methanol maser observations, to derive a parallax distance of $3.27 \pm 0.49$ kpc. 
The complexity of the region is revealed by interferometric radio observations that show that the molecular cloud contains a PDR region \citep{roshi2005} with five HII regions, of which \gtcdz\ is the brightest one  \citep{onello1994}. In addition, observations with higher angular resolution reveal a cometary morphology for  \gtcdz\ with the tip pointing to the north-east \citep{wood1989,kurtz1994}.  
Strikingly, low resolution maps of this region at cm wavelengths show large scale emission oriented to the north-west \citep{roshi2005}.

As seen in Figure \ref{g35_20}, our observations recover the north-eastern tip of the region and show the compact source G35.20-VLA1, reported for the first time. This source is located near the center of the UCHII region and $\lesssim$ 1$''$ south-west of the brightest emitting region. Thus, the source is clearly detached from this part of the UCHII region but still embedded in the extended cometary emission. Among our sample of compact sources, G35.20-VLA1 has the largest emission measure and ionized gas density ($\sim3 \times 10^9$ cm$^{-6}$pc$^{-1}$ and $\sim2 \times 10^5$ \cmt, respectively, see Table \ref{physical_parameters}).



\item[-] \emph{\gssuv.} The { UCHII region} appears unresolved in the maps presented by \citet{kurtz1994} with an angular resolution of $\lesssim$1 \arcsec. We derive a spectral index of $-0.45 \pm 0.05$ using the 8 GHz bandwidth of our observations, which is indicative of non-thermal emission. We found no evidence for circular polarization. From our higher resolution data we can obtain a very small and uncertain size. Given the poor $uv$ coverage of our observations, we do not attribute significance to this deconvolved size.  Therefore, we cannot discard a stellar corona origin for the emission of this source. Possibly, such emission is associated to a low mass stellar companion of the exciting star of the region.

 \item[-] \emph{\gsstv.} This region belongs to the S106 HII region located at a distance of 1 kpc \citep{eiroa1979,churchwell1982}.  The region shows a bipolar morphology, possibly indicative of a powerful wind \citep{felli1984}. There is some controversy about the exciting object of the region, but most of the studies point to an early B-type star \citep[e.g.][]{bally1983, kurtz1994}. Despite the complexity of the region, \citet{kurtz1994} only detected the compact core probably associated with the exciting star. Our higher angular resolution observations show similar results with a deconvolved compact core size of $\sim100 \times 50$ AU. 
  
 Because of the high signal to noise ratio of our detection for this source, we could derive the spectral index within the 8 GHz of bandwidth of our observations. Our result, $0.75 \pm 0.08$, is consistent with an ionized outflow. Besides, our data show marginal circular polarization over the source at a level of $0.5 \pm 0.1\%$. In order to confirm or reject this result, we analyzed the data from project AF362 taken on 1999 July 6 at X-band and
set an upper limit of $\leq 1.1\%$ (3-sigma upper limit) for the circular polarization. We
conclude that
there is no definitive evidence of circular polarization for this source. 

 

\end{itemize}

\subsubsection{The nature of Type I sources} 

The close association with the centimeter emission peak and with a possible IR counterpart suggest that the sources of Type I are associated with the exciting star of the region. However, for \gdvdn\ and \gtcdz, \citet{kurtz1994} estimated a spectral type of O9 for the ionizing star. Such a star delivers $\sim5 \times 10^{48}$ s$^{-1}$ ionizing photons  \citep{panagia1973} and, as seen in column 7 of Table \ref{physical_parameters}, the ionizing flux required to maintain the free-free emission of G28.29-VLA1 and G35.20-VLA1 is $\sim5 \times 10^{45}$ s$^{-1}$. Therefore, these compact sources probably do not harbor the exciting star of the region unless a considerable leakage of the UV photon flux occurs in the source (see the analysis of G35.20-VLA1 below).

The scenario where these sources are very compact HII regions surrounding an early B type star embedded in the UCHII region is unlikely: such a region would be dynamically unstable and expand away with a lifetime too short to be observed. However, a mechanism to maintain these very compact photo-ionized regions observable for a longer time could be present. Given their small size, a possibility is that they are confined within the region where the material is gravitationally trapped. This could be the case of G35.20-VLA1 and G76.38-VLA1, whose sizes are smaller than twice the gravitational radius for an O9 ($\sim150$ AU) and B1 ($\sim125$ AU) stars, which are ionizing \gtcdz\ and \gsstv, respectively. Thus, if these sources harbor the exciting source of the region they could retain a significant amount of ionized material, especially G35.20-VLA1, whose electron density and emission measure are an order of magnitude above those derived for the rest of the compact sources.  Also, G76.38-VLA1 and the 2MASS source positions are an excellent match and, in addition, the required ionizing photon flux for this source is consistent with a B1 star.

 

Possibly, these objects are not fully ionized but, instead, they may have an inner neutral component. This neutral component would be continuously photo-evaporated, either internally or externally, by the ionizing star of the region. Such a scenario for a photo-evaporating accretion disk was proposed by \citet{hollenback1994}. Otherwise, they could be dense neutral globules surrounded by an externally ionized envelope \citep{garay1987}. These objects are probably small scale structures inherent of molecular clouds \citep[e.g.][]{pauls1983,morata2003} that were engulfed by the ionizing front. 

We propose that G28.29-VLA1 is externally ionized by the O type star exciting the region because its size is larger than twice the gravitational radius of the exciting star. 
In this case, the number of ionizing photons causing the ionization is set by the stellar type and the solid angle of G28.20-VLA1 as seen from the exciting star. As discussed above, the required UV photon flux necessary to ionize G28.29-VLA1 is $4.6 \times 10^{45}$ s$^{-1}$. That means that the number of ionizing photons impinging G28.20-VLA1 is a factor of 10$^3$ lower than the total provided by the exciting star. Thus, taking into account the distance of \gdvdn\ \citep[3.3 kpc,][]{solomon1987} and the size of G28.29-VLA1 (300 AU), the exciting star must be located at 2600 AU from the compact source in order to obtain a geometrical dilution of 10$^3$. Considering that the dust opacity affects the UV photons in the region, the separation between G28.29-VLA1 and the ionizing star must be taken as an upper limit. Similarly, considering possible projection effects in the map, the observed angular distance can be smaller than that corresponding to the separation between the exciting source and G28.29-VLA1. According to this, our estimated value for the angular separation, $\lesssim0.8$ arcsec, is in good agreement with the separation of the IR and radio sources as seen in the corresponding panel of Fig. \ref{compact_a}.

Since G35.20-VLA1 and G76.38-VLA1 possibly harbor the exciting source of the region, we propose that they are internally ionized. From their deconvolution, these sources appear to be elongated with a shape reminiscent of a partially edge-on disk-like morphology. In this scenario, the non detection of any IR counterpart of the exciting star in G35.20-VLA1 could be explained if it is surrounded by the edge-on disk and efficiently obscured. Moreover, in the case of G76.38-VLA1, the source is elongated perpendicular to the outflow found by \citet{felli1984}. This invokes the photo-evaporating disk picture around a massive star proposed by \citet{hollenback1994}. In this picture, the ionized material forms a static atmosphere above the disk for radii smaller than the gravitational radius. If the stellar wind is not excessively strong, the static atmosphere is preserved and produces the observed free-free emission. Since HCHII regions are considered to harbor only a single star or small multiple system, the true nature of these regions could be explained as static (or blowing) ionized atmospheres of these photo- evaporating disks around massive stars. However, the shape and extension of these static atmospheres depends on the stellar wind properties, which for the exciting star of these regions are poorly constrained. The study of these sources based on a possible photo-evaporating disk is beyond the scope of this paper and we propose this interpretation as tentative.

\subsection{Sources of Type II}

Sources of Type II are not located within the boundaries of their associated UCHII region.
Although the sources of this category probably belong to the main region, their location suggests that they are unrelated with the dense gas of the UCHII region. Below we give some comments on individual sources and a speculative discussion on the possible nature of sources of Type II.   

\begin{figure}[thbp]
\vspace{-3.6cm} 
\resizebox{1.15\textwidth}{!}{\includegraphics[angle=0]{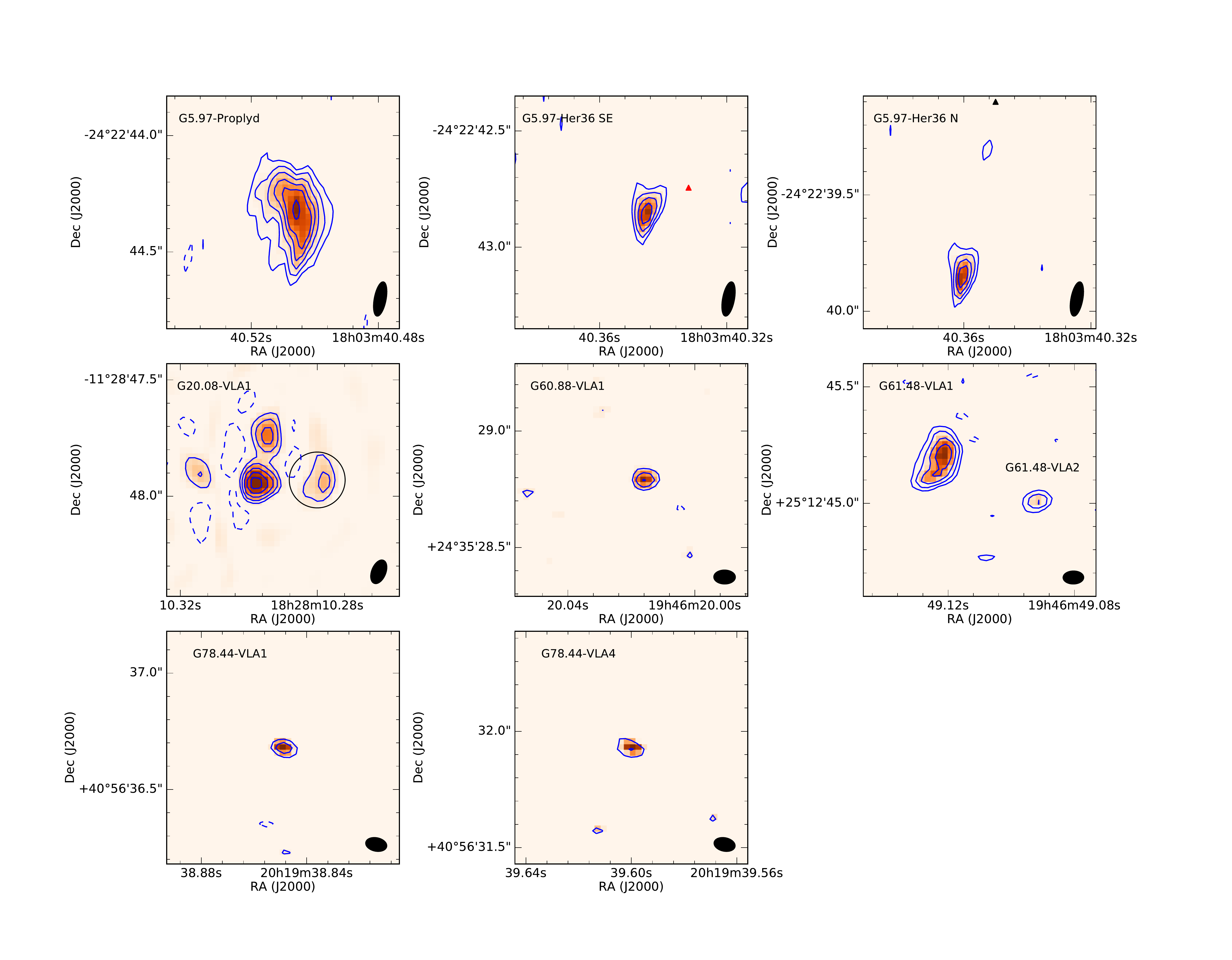}}
\caption{
Close-up continuum maps (color scale and contours) of compact sources of Type II  (i.e. detached from the UCHII region) constructed with uniform weighting and removing the short spacings of the visibility dataset for selected regions, except for the G20.08-VLA 1 map that was obtained with natural weighting. The corresponding largest angular structure mapped is shown in Column 4 of Table \ref{maps}. For all panels, levels are  -3, 3, 6, 10, 15, 20, and 30 times the $rms$ noise shown in column 6 of Table \ref{maps}. The synthesized beam is shown in the bottom right corner. The circle in the G20.08-VLA 1 panel shows the corresponding compact source. 
The 2MASS (PSC) position is marked as a red triangle in the G5.97-Her36 SE panel. The black triangle seen in the G5.97-Her36 N panel shows the position of G5.97-Her36 B.} \label{compact_b}
\end{figure}

\begin{figure}[thbp]
\vspace{-3.6cm} 
\resizebox{1.2\textwidth}{!}{\includegraphics[angle=0]{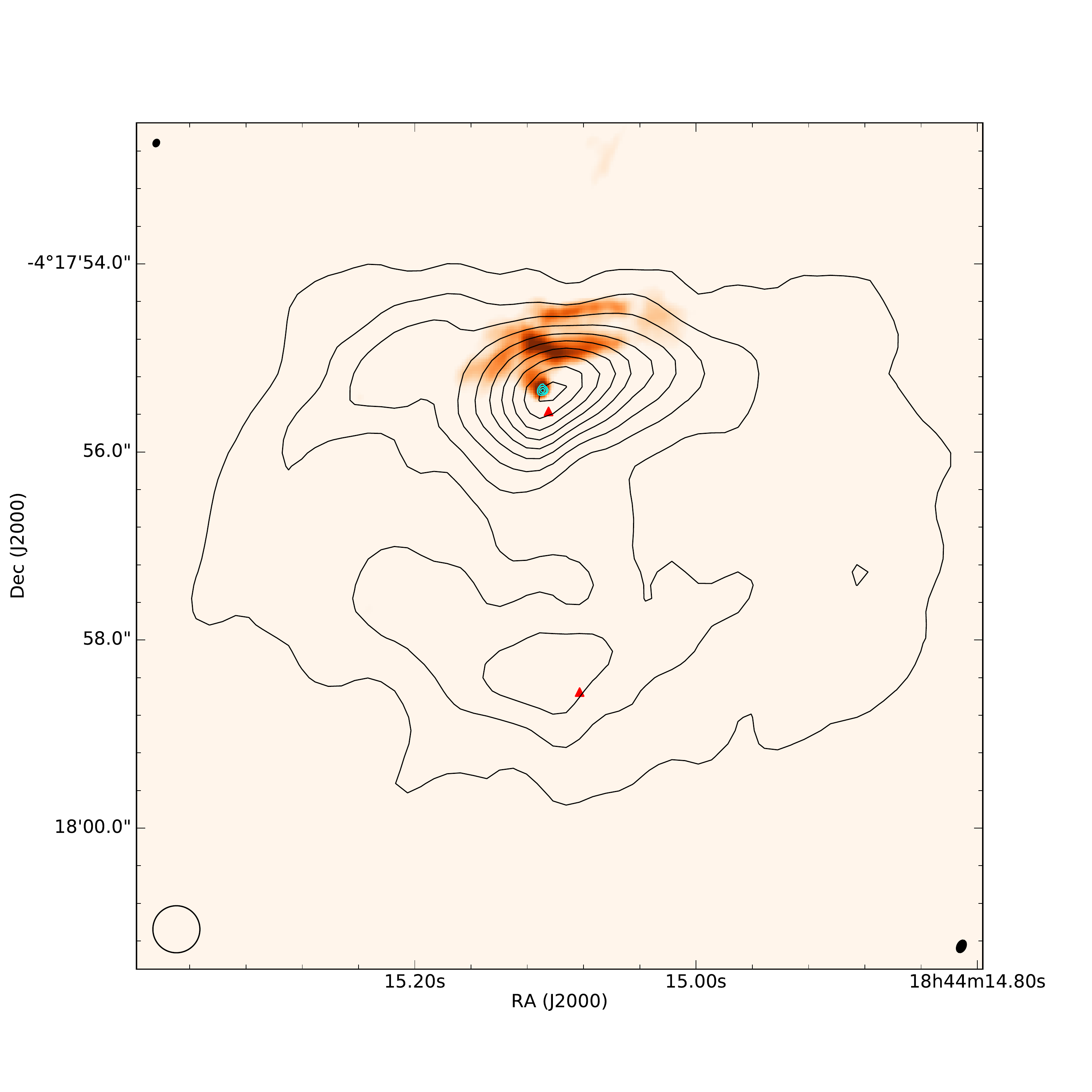}}
\figcaption{
Map of the 2 cm continuum emission of the \gdvdn\ region of \citet{kurtz1994} (black contours) overlaid to the 1.3 cm continuum emission (this work). For the 1.3 cm emission we show the map obtained with natural weighting (color scale) and the map obtained with uniform weighting without short spacings (cyan contours). Black contours are the same as in the corresponding figure of \citet{kurtz1994}.  
The cyan contours are 0.5, 0.7 and 0.9 times the peak value of the corresponding map (1 m\jpb). The synthesized beam of the 1.3 cm natural weighted map is shown in the bottom right corner and that of the uniform weighted map is shown in the top left corner. The approximate beam of the 2 cm observations is shown in the bottom left corner as an open circle. The red triangles shows the position of the 2MASS sources. \label{g28_29}}
 \end{figure}

\begin{figure}[thbp]
\vspace{-3.6cm} 
\resizebox{1.2\textwidth}{!}{\includegraphics[angle=0]{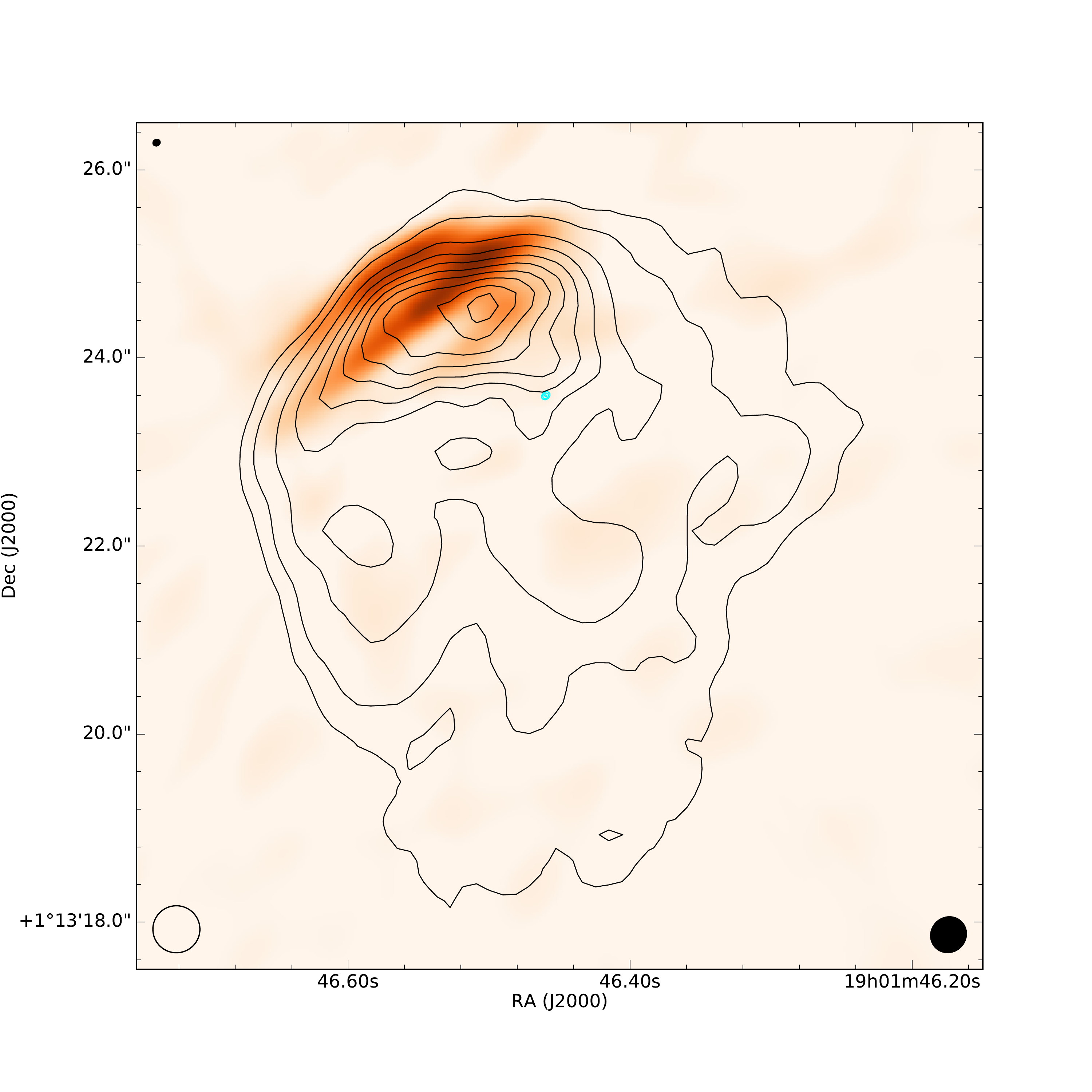}}
\caption{
Map of the 2 cm continuum emission of the \gtcdz\ region of \citet{kurtz1994} (black contours) overlaid to the 1.3 cm continuum emission. For the 1.3 cm emission we show the map obtained with natural weighting (color scale) and the map obtained with uniform weighting without short spacings (cyan contours). Black contours are the same as in the corresponding figure of \citet{kurtz1994}.  
The cyan contours are 0.5, 0.7 and 0.9 times the peak value of the corresponding map (3.5 m\jpb). The synthesized beam of the 1.3 cm natural weighted map with 0.35 arcsec of tapering is shown in the bottom right corner and that of the uniform weighted map is shown in the top left corner. The approximate beam of the 2 cm observations is shown in the bottom left corner as an open circle. \label{g35_20}
 }
\end{figure}

\begin{itemize}

\item[-] \emph{\gcns.} This region lies in the core of an extended HII region, the Lagoon Nebula or M 8, located at a distance of 1.3 kpc \citep{arias2006}.  
The region is ionized by the UV photons of an O7 star named Herschel 36 \citep[][hereafter, Her 36]{woolf1961}. About 4000 AU southeast from Her 36 there is a radio source identified with \gcns\ that was first associated with an UCHII region. More recently, however, \citet{stecklum1998} interpreted this source as a proplyd (hereafter G5.97). Besides, \citet{goto2006} reported the presence of a close intermediate-mass stellar companion of Her 36 traced by 2 cm emission (Her 36 SE). Using data of the present paper, \citet{masque2014} confirmed the proplyd nature of G5.97 finding hints of non-thermal emission from this object and, at the same time, detected radio emission towards the Her 36 multiple system.         

The map of Figure \ref{general} shows the presence of three compact sources in the \gcns\ region. The eastern source corresponds to the G5.97 proplyd and is the largest object in angular size of the list of compact sources of Table \ref{sources}. Its proplyd nature is suggested by its cometary shape with the tip pointing to Her 36 (see Fig. \ref{compact_b}). We derived an emitting size of $330 \times 160$ AU that is in good agreement with the radius of 160 AU derived in \citet{stecklum1998}. The radio emission of Her 36 appears
2\rlap{.}$''$7 northwest of  the G5.97 proplyd as an unresolved source. \citet{masque2014} associated this radio emission to a region where the winds of Her 36 and Her 36 SE collide, owing to the position of the radio peak between  the two stellar components. The 2MASS position seen in the G5.97-Her 36 SE panel of Fig. \ref{compact_b} is slightly displaced to the north-west of the radio emission and it is possibly coincident with the Her 36 main component.

North of Her 36 there is another unresolved radio source with a flux density of 1 mJy (that we call Her 36N). Consulting the literature we find that \citet{allen1986} reports a northern IR companion (Her 36B) located 3.6 arcsec from the Her 36 main component. Although the association of the radio source with Her 36B is tempting, there is a small but significant offset of $\sim0.8$ arcsec between the position of these sources. Instead, the Her 36N position has an excellent match with a 2.2 $\mu$m source seen close to Her 36B in the Figure 1 of \citet{goto2006}.


 \item[-] \emph{\gdzzv.} The \gdzzv\ region is embedded in a molecular cloud \citep{turner1979, plume1992}  where massive stars are forming as confirmed by maser observations \citep{ho1983, hofner1996,walsh1998}. \citet{galvanmadrid2009} carried out a detailed study of the kinematics of the region through molecular and recombination line observations and found that an accretion flow occurs at multiple scales. Using the new Bayesian distance calculator from \citet{reid2016} and a $v_{\rm lsr}\simeq45$ \kms \citep{argon2000}, we obtained that the near and far kinematic distances to G20.08-0.14 are $3.36\pm0.18$ kpc, and $12.30\pm0.20$ kpc, with probabilities of 0.85 and 0.15, respectively. Given that the probability of the nearest kinematic distance is so much higher, we will adopt it.


  The region is composed of three components labelled A, B and C, where region A is the brightest and region C the most extended \citep{wood1989}. Our observations show the regions A and B, and resolve out the C region.  Region A is the brightest one as expected and region B shows a ring-like morphology, maybe representing a signpost of an expanding bubble interacting with the medium. In any case, as we are interested in small components, we focus our attention on the compact source blended to the western side of the A region (G20.08-VLA1). Some properties of  G20.08-VLA1 differ from the rest of the compact sources. First, this source has a significantly larger size ($\sim500 \times 200$ AU), even though its determination is uncertain because it appears blended with the A region. Second, G20.08-VLA1 can be associated with hot core molecular emission (OCS and CH$_3$CN) mapped by \citet{galvanmadrid2009}.




\item[-] \emph{\gszvv.} This region, also known as Sh 87, is an active star-forming region
located at a distance of 2.1~kpc \citep{clemens1985}. This region, catalogued
as an optical H~II nebula, is also associated with the infrared source
IRAS 19442+2427 and has been studied by several authors at different
wavelengths \citep[e.g.][]{barsony1989,xue2008}. In particular,
an UCHII region was detected at 2 and 3.6~cm by  \citet{kurtz1994}, with an embedded B0.5 ZAMS star. In addition, Sh 87 is one of the few examples where the proposed mechanism to trigger star
formation as a cloud-cloud collision appears to be observed \citep{xue2008}.

G60.88-VLA1 is located 2$\arcsec$ west of the UCHII region and appears unresolved. This source is coincident within 1.5 arcsec with a YSO reported by \citet{campbell1989}. According to the position accuracy of their observations (2 arcsec), the association of both objects is possible. Moreover, there are several IR sources in the field suggesting the presence of a cluster of young objects. All these features suggest a pre-main sequence stellar nature for G60.88-VLA1.

\item[-] \emph{\gsuqv A.} This object is an active star-forming region, located in the emission nebula Sh 2-88B.
Its distance is controversial, but assuming 2.5~kpc, \citet{evans1981} estimated
a far-infrared luminosity of $2.8\times10^5$~L$_{\sun}$. However, in the present work we will use the distance adopted in \citet{kurtz1994} of 2.0 kpc. G61.48$+$0.09A is composed of an extended
cometary (B1) and an UC (B2) HII regions \citep{felli1981}. \citet{deharveng2000} associated B1
with a cluster of high-mass stars, with an O8.5V-O9.5V star as the dominant exciting
source, while B2 appears as an UC HII region with an exciting star of spectral type
later than B0.5V.
  
 Furthermore, in \citet{wood1989}, G61.48$+$0.09A appears as a compact northern source with some extended emission to the south. Our observations reveal the northern source as a bow shock structure pointing to the south-west and two additional  compact sources embedded in the diffuse southern emission (see the map of the region in \citealt{wood1989}). One of them is clearly resolved (G61.48-VLA1) while for the other the deconvolved size is poorly determined (G61.48-VLA2) implying a physical size of $\leq100$ AU. None of the compact sources have known counterparts at other wavelengths possibly due to obscuration, as the region is embedded in a large dense cloud. 


\item[-] \emph{\gsvqq.} The \gsvqq\ region is associated with IRAS 20178$+$4046 and harbors a cluster of young stars detected at IR wavelengths \citep{tej2007}. Interferometric radio observations show that the UCHII region has a cometary morphology with the tip pointing to the North-West \citep{kurtz1994}. Our A configuration observations, however, miss all the extended emission and only two compact sources appear in the map. One of these sources, G78.44-VLA4,  can be associated with an IR source previously detected by \citet[][their source 2]{tej2007} and with a compact radio source \citep[][their VLA4, we will adopt the same name]{neria2010}. These latter authors detected a group of compact radio sources that they attributed to the presence of a cluster of pre-main sequence stars. 

G78.44-VLA1 is unresolved and has no known counterparts. We look for circular polarization toward this source with null results. Contrary to G78.44-VLA4, G78.44-VLA1 appears far away from the extended centimeter emission.


 \end{itemize}

\subsubsection{The nature of Type II sources}

    Only three out of the nine sources of Type II are resolved (G5.97-Proplyd, G20.08.VLA1 and G61.48-VLA1). G20.08-VLA1 is the largest source of our sample and seems to have a well defined nature: it probably corresponds to a hot core with an embedded massive object that has begun to ionize the internal part of the core. The molecular emission would come from a neutral shell surrounding the ionized region, where the centimeter emission is produced. The ionized region is expected to expand at the speed of sound that, for the derived size of G20.08-VLA1 ($\sim500$ AU), yields an estimated age of about 200 yr. Similarly, the measured size in the map for the bubble corresponding to the region C ($\sim3000$ AU) yields $\sim10^3$ yr of lifetime. The youth of both objects makes them unlikely to be observed but not impossible given the crowded region where they are situated. The fact that G20.08-VLA1 appears attached to other ionized regions of \gdzzv\ but is not embedded within any of them, suggests that it is part of a cluster of UC (or HC)HII regions. In this case, the improvement of our data with respect to previous observations allows us to detect smaller mass and/or younger members of the cluster.      
     
  G5.97-Proplyd and G61.48-VLA1 have sizes slightly larger than the resolved sources of Type I. Their emission measure and density are below those typical for an HCHII region (see Table \ref{physical_parameters}). Their cometary morphology is reminiscent of proplyds. While this is confirmed for G5.97-Proplyd \citep{stecklum1998,masque2014}, from the present data we cannot discriminate between a proplyd or globule nature for G61.48-VLA1, unless an IR counterpart indicating the presence of a stellar component is detected.    
    
   Except for G5.97-Her 36SE, whose emission is interpreted as synchrotron radiation produced in a wind collision region in the close binary system of Her 36, it is not possible to distinguish between various scenarios to explain the nature of the rest of the unresolved compact sources of Type II. The possibility of detecting a protostellar wind is unlikely: the unresolved compact sources presented in this paper have flux densities  of a few mJy. Assuming a flux density of 1 mJy at 1.3 cm, adopting a terminal wind velocity of 2700 \kms\ \citep{dzib2013b} and following the formulation of \citet{panagia1975}, we obtain a mass loss rate of $\sim10^{-5}$ \msun\ yr$^{-1}$. This would imply the improbable presence of an early O type star in the region. Alternatively, taking into account wind collimation, a fully ionized jet would provide a larger emitting flux. However, a jet like structure would show a resolved size at least along one axis.
  
 Three unresolved sources have IR counterparts (G5.97-Her36 N, G60.88-VLA1 and G78.44-VLA4). The IR emission could be due to source heating maybe caused by the presence of a stellar object. The rest of the unresolved sources (G61.48-VLA2 and G78.44-VLA1) have no IR association even though we cannot discard that they are located in regions of high extinction. One possibility is that the centimeter emission of the unresolved sources have a stellar corona origin. For instance, previous works suggest a pre-main sequence star for G60.88-VLA1, owing to the detection of an IR cluster in the region \citep{campbell1989}. Otherwise, the centimeter emission could be produced in  a small HII region trapped by a B type star present in the region, since the upper limits for the source size ($\leq300$ AU, even though the real size of the compact source is much smaller) are larger than twice the gravitational radius of such a star. This could explain the location of G61.48-VLA1 next to G61.48-VLA2 with the tip pointing to G61.48-VLA2, which makes the latter a good candidate to photo-evaporate G61.48-VLA1 (see Fig. \ref{compact_b}). This possibility is also proposed for G78.44-VLA4 \citep{neria2010}. Observations at another frequency in order to derive spectral indices are required to discern between the two proposed scenarios. In any case, the presence of possible young stellar objects in the region would be a manifestation of the important stellar multiplicity in the vicinity of massive stars.

\subsection{Lifetime of UCHII regions}

Regardless of the nature of these sources, the frequent presence of photo-evaporating objects in UCHII regions could provide a solution to the so-called \emph{lifetime} problem of these regions. The photo-evaporated material could replenish continuously the expanding plasma maintaining large emission measures in a compact area \citep{hollenback1994,lugo2004}.  As a result, while the UCHII region is expanding dynamically,  
it would be observable for a significantly longer time than it would do in absence of \emph{replenishment}. This possibility also applies for the UCHII regions with null detection of compact sources belonging to Type I, since compact components can be present in the region but be missed by our filtering technique. This is more probable if the compact sources are coincident with the brightest part of the UCHII region, specially for the resolved ones. Observations with a better $uv$ coverage are required to prevent this effect.

In the case of external photo-evaporation, the photon flux incident on the neutral condensation, $J_0 = \dot{N}_{star} / 4 \pi d^2$, with $d$ being the projected distance to the exciting source that produce  $\dot{N}_{star}$ ionizing photons,  is reduced as $J = 2J_0/ [ 1 + \sqrt{1 + \alpha \dot{N}_{star} R/3\pi d^2c^2 } ]$   \citep{spitzer1978} where $J$ is the photon flux at the ionization front of the condensation, $c$ is the speed of sound in the medium ($\sim10^6$ cm s$^{-1}$),  $\alpha $ is the recombination coefficient ($3 \times 10^{-13}$ cm$^3$ s$^{-1}$) and $R$, the radius of the condensation. Then, equating the rate at which gas is ionized at the ionization front with the change of neutral mass per unit time, we get $\dot{M} = \pi R^2 J \mu m_\mathrm{H}$, where $\mu$ is the mean molecular weight per particle, adopted to be 2.3, and $m_\mathrm{H}$ is the hydrogen mass.

In the likely scenario where the structure of these photo-evaporating objects is composed by a neutral component with a thin ionized envelope, we can study the case of G28.29-VLA1, which is a Type I source being externally photo-ionized. Adopting as the radius of the object half the derived size of the compact source (150 AU, see Table \ref{physical_parameters}) and a distance of 825 AU to the ionizing source estimated from the maps, we obtain a mass depletion rate due to photo-evaporation of $3 \times 10^{-6}$ \msun\ yr$^{-1}$. This rate is about an order of magnitude larger than those derived for the photo-evaporating objects in the ONC \citep[$10^{-6}-10^{-7}$ \msun\ yr$^{-1}$,][]{garay1987,churchwell1987}. Such amount of photo-evaporating material could create a region around G28.29-VLA1 with an emission measure large enough to be observable as an UCHII region. 

Assuming a steady photoionization rate, a lifetime can be estimated from $M/\dot{M}$, where $M$ is the mass of the condensation.
\citep{garay1987,churchwell1987} estimated masses for the Orion photo-evaporating objects of the order of 0.1 \msun. Given the slightly larger size of G28.29-VLA1 with respect to the Orion sources ($\sim300$ AU vs. 100-200 AU, see Table 5 of \citealt{churchwell1987}) and the fact that the former is embedded in a higher pressure ambient, we can adopt this mass as a lower limit for the G28.29-VLA1 mass. Our derived lifetime of $\geq3 \times 10^4$ yr is larger than the typical lifetime of UCHII regions estimated from free expansion.
 
Similarly, a photo-evaporating disk around the exciting star, as proposed for G35.20-VLA1 and G76.38-VLA1, could provide enough ionized material to replenish the UCHII region. \citet{hollenback1994} estimated that disk masses of a few \msun\ can expand considerably the lifetime of the region. Although from the present data we cannot infer the mass of the disks, their estimated radius (50-120 AU) are somewhat smaller but of the order of other disks found around massive stars \citep[e.g. IRAS 18162$-$2048: 200 AU,][]{carrascogonzalez2012} that are known to have at least a few solar masses \citep{fernandezlopez2011b,carrascogonzalez2012}. According to \citet{hollenback1994}, such a disk would supply ionized material to the region for $\gtrsim 10^5$ yr.

 The same calculation of mass depletion rate shown above can be applied to photo-evaporating objects representative of Type II such as G5.97 Proplyd or G61.48-VLA1. For G5.97 Proplyd, which is being photoionized by an O7 spectral type star (Her 36) situated 3500 AU away in projection, we derive a depletion rate of $\sim10^{-6}$ \msun\ for a 150 AU radius (see column 8 of Table \ref{physical_parameters}). This value is higher than the value derived by \citet[][$7 \times 10^{-7}$ \msun]{stecklum1998} possibly because they adopted a larger distance to the region (1.8 kpc). Performing the same analysis for G61.48-VLA1, which has a radius of 150 AU and is separated by 1000 AU from its supposed ionizing star (G61.48-VLA2), we obtain a mass depletion rate of  $\leq 10^{-6}$  \msun\ yr$^{-1}$ if we assume that the latter object is a B type star ($\dot{N}_{star}  \leq 10^{48}$ s$^{-1}$). These rates are similar to those found for photo-evaporating objects in the ONC and lower than those of the sources of Type I. The similarity between G5.97 Proplyd and G61.48-VLA1 with the Orion photo-evaporating objects suggests similar lifetimes (10$^6$ yr). Therefore, the photo-evaporating sources of Type I are statistically younger than the photo-evaporating sources of Type II since the latter have lower mass depletion rates ($10^{-6}-10^{-7}$  \msun\ yr$^{-1}$) and can last an order of magnitude longer ($\sim10^6$ vs. $\sim10^5$ yr).

\section{Discussion and Conclusions}

From the last column of Table \ref{sources} we can assess the high occurrence of UCHII regions with associated compact components: 9 out of the 12 UCHII regions surveyed have compact sources associated with them. In total, we detected 13 such sources. Nevertheless, the expected number of compact sources associated to a given UCHII region is low: we normally find one compact source per UCHII region and only in three cases did we find additional compact sources associated with the same region. Invoking the statistical arguments discussed in Sect. 3, the likelihood of detecting background sources in our field of view is very low and we consider that these compact sources are indeed associated with the corresponding UCHII region. In addition, most of them present free-free emission indicating that they are (at least partially) ionized. This may add evidence that they are physically related to the region.

  

We classify these compact sources as Type I if they are located within the boundaries of their associated UCHII region, or Type II if they are associated with an UCHII region but not detected directly toward said region.  As seen in the previous section, Type I sources include G28.29-VLA1, G35.20-VLA1, G76.38-VLA1 and G76.18-VLA1, even though the latter is too far away to determine properly its features. As seen in column 4 of Table \ref{sources}, these sources can be spatially resolved with sizes typically in the range $\sim100-300$ AU, discarding a stellar corona origin for their emission. Although our observations are insufficient to fully elucidate the intrinsic nature of these sources, they seem to be photo-evaporating objects such as protostellar disks or globules with a neutral component. However, we infer important differences between them: while G28.29-VLA1 is probably externally photo-ionized, the rest of the sources of this group appear to harbor the exciting star of the region. As seen in the previous section, the derived photo-evaporating rates for Type I sources are clearly higher than that of photo-evaporating proplyds and globules in the ONC. This might produce observational features such as UCHII regions, as a result of the considerable amount of ionized material flowing around the compact source.


On the other hand, sources of Type II includes G5.97-Proplyd, G5.97-Her36, G5.97-Her 36N, G20.08-VLA1, G60.88-VLA1, G61.48-VLA1, G61.48-VLA2, G78.44-VLA1 and G78.44-VLA4. A prime example is provided by \gszvv, where G60.88-VLA1 is located far away from the cometary structure of the UCHII region. G20.08-VLA1 has likely a distinct nature and its discussion is beyond the scope of this paper. The high detection rate for these sources could be a consequence of the high population of objects associated to massive star forming regions. In the most recent survey of radio sources over the ONC, 556 sources were detected \citep{forbrich2016}. This high detection rate was due to the depth of the observations, which for a nearby region such as Orion, provided an unprecedented threshold for radio source detection. Among the regions observed in this work, \gcns\ has a distinctive extended diffuse appearance and, hence, it is the one that most resembles the ONC. If Orion was located at the distance of \gcns, the flux density of their compact sources would be roughly a factor of 10 lower than the values reported in \citet{forbrich2016}. Adopting the same detection threshold of signal to noise ratio greater than 5, the $rms$ value shown in the column 6 of Table \ref{maps} yields to 0.25 m\jpb. Only $\sim10$ sources from the \citet{forbrich2016} catalog would be above this level if their fluxes were scaled to that distance. The difference of field of view between their C band observations and our Ku band observations (the latter is about 5 times smaller) is approximately balanced by the difference of distance between Orion and M8. Thus, in terms of lineal size we are mapping a similar region than that of \citet{forbrich2016}.  The number of detected radio sources in M8 by us (3) is somewhat below the number expected for a ONC-like region ($\sim10$). However, given the uncertainties of this analysis we cannot discard that M 8 has characteristics similar to the ONC in terms of radio source population. In this picture, many more compact sources remain to be discovered in M8; they are currently below our detection threshold. Very deep interferometric radio observations over the M8 region would assess the existence of a similar population of radio sources as in the ONC.


The rest of the regions that have associated sources of Type II also could harbor a rich variety of radio sources as that of the ONC. Indeed, some compact sources of Type II are unresolved and some of them are related with already \emph{visible} stellar objects such as PMS stars. This would correspond to a later evolutionary stage where the stellar object has already cleared out the envelope.

As a trend, \citet{forbrich2016} found that the brightest sources of the ONC are predominantly thermal. This was attributed to the larger emitting volume of the ionized gas around thermal sources with respect to non-thermal sources, where the emitting volume is restricted to the surface of the stellar corona. 
Indeed, the largest and  intrinsically brightest resolved sources of Type II detected here (G20.08-VLA1, G5.97 Proplyd and G61.48-VLA1) are likely thermal nature and, with the exception of G20.08-VLA1, they are photo-evaporating objects similar to the sources of Type I. However, their photo-ionizing rate is about an order of magnitude lower, comparable  to those of ONC objects. Thus, 
the flowing plasma from the photo-evaporating objects of Type II is expected to be more diffuse than that of the objects of Type I, and possibly more extended as Type II sources are statistically older than sources of  Type I. Therefore, the associated centimeter emission of the photo-evaporated material of the Type II objects would be weaker and more extended than that of the objects of Type I, and thus easily filtered out by an interferometer.

From the present data we cannot establish an unambiguous nature for the newly detected compact sources.
However, we assess some of their observational properties and their implications in the UCHII region evolution, which are listed below:

\begin{enumerate}

\item  We found 13 compact sources among our observed 12 regions, indicating a high occurrence of such sources associated with UCHII regions or lying close to them. Based on this, we categorize the sources into two distinctive groups of compact sources (Type I and Type II).

\item Type I corresponds to sources coincident with the bright free-free emitting area and are possibly embedded in the UCHII region. They are plausibly photo-evaporating objects with mass depletion rates of 10$^{-5}-10^{-6}$ \msun\ yr$^{-1}$ and lifetimes of 10$^4-10^5$ yr. These objects would be extremely close to the ionizing massive stars and still would have some reservoir of mass due to their youth. They would show observational features such as UCHII regions lasting for $\geq10^4$ yr, as a result of the considerable amount of ionized material flowing around the compact source. This interpretation provides a solution to the so-called lifetime problem for UCHII regions.

\item The sources of Type II appear unrelated to the dense gas of the UCHII region. Some of these sources could be photo-evaporating objects with mass depletion rates of $\leq10^{-6}$ \msun\ yr$^{-1}$.
This flowing material is too diffuse to create an observable UCHII region. Other sources of Type II appear to be unresolved and possibly correspond to stellar objects. All these characteristics suggest that Type II sources correspond to a more evolved population compared to Type I sources.

\item There are some similarities between Type II sources and those of ONC radio sources. In particular, both populations have a rich variety of objects. Moreover, this variety includes photo-evaporating objects, such as globules and proplyds, which have mass depletion rates similar to those of ONC photo-evaporating objects. Possibly, the compact sources detected here are the tip of the iceberg of a larger and varied population of radio sources similar to that of the ONC.

\end{enumerate}

%


 \acknowledgments

We thank an anonymous referee for his/her constructive report. LFR and LL acknowledges the support of DGAPA, UNAM and CONACyT, México.

\bibliography{HIIregions_paper}

\begin{landscape}
\begin{table}[ht]
\footnotesize
\begin{center}
\vspace{-1.7cm}
\caption{Summary of the VLA observations\label{regions_observed} }
\begin{tabular}{p{2.5cm}ccccc}
\hline
\hline
& \multicolumn{2}{c}{Pointing Center} &  & Bootstrapped Flux Density$^{\mathrm{a}}$  & Fitted Spectral Index\\
Source   &  $\alpha$~(J2000) & $\delta$~(J2000) & Gain Calibrator & (Jy) &  \\
\hline
G5.89-0.39 & $18^\mathrm{h}00^\mathrm{m}30\rlap{.}^\mathrm{s}387$  & $-24^{\circ}04'00\rlap{.}''12$ & J1745-2900 & $0.99 \pm 0.01$  & $ 0.011 \pm 0.02 $ \\
G5.97-1.17 & $18^\mathrm{h}03^\mathrm{m}40\rlap{.}^\mathrm{s}504$  & $-24^{\circ}22'44\rlap{.}''40$ &  & &  \\
G19.61-0.23 & $18^\mathrm{h}27^\mathrm{m}38\rlap{.}^\mathrm{s}117$  & $-11^{\circ}56'39\rlap{.}''48$ &  J1832-1035  & $1.043 \pm 0.005 $ & $ -0.718 \pm 0.002 $ \\
G20.08-0.14 & $18^\mathrm{h}28^\mathrm{m}10\rlap{.}^\mathrm{s}280$  & $-11^{\circ}28'47\rlap{.}''13$ &  & & \\
G28.29-0.36 & $18^\mathrm{h}44^\mathrm{m}15\rlap{.}^\mathrm{s}097$  & $-04^{\circ}17'55\rlap{.}''29$ &  J1851-0035  & $0.803 \pm 0.004 $ & $ -0.200 \pm 0.006 $ \\
G35.20-1.74 & $19^\mathrm{h}01^\mathrm{m}46\rlap{.}^\mathrm{s}490$  & $01^{\circ}13'24\rlap{.}''65$ &  & & \\
G60.88+0.13 & $19^\mathrm{h}46^\mathrm{m}20\rlap{.}^\mathrm{s}130$ & $24^{\circ}35'29\rlap{.}''39$ & J1931+2243 & $0.517 \pm 0.004$ & $-0.171 \pm 0.001$ \\
G61.48+0.09A & $19^\mathrm{h}46^\mathrm{m}49\rlap{.}^\mathrm{s}202$ & $25^{\circ}12'48\rlap{.}''05$ &  & & \\
G76.18+0.13 & $20^\mathrm{h}23^\mathrm{m}45\rlap{.}^\mathrm{s}733$ & $37^{\circ}38'32\rlap{.}''38$ & J2015+3710 & $3.86 \pm 0.01$  & $-0.006 \pm 0.001$ \\
G76.38-0.62 & $ 20^\mathrm{h}27^\mathrm{m}26\rlap{.}^\mathrm{s}773$ & $37^{\circ}22'48\rlap{.}''01$ & & & \\ 
G78.44+2.66 & $ 20^\mathrm{h}19^\mathrm{m}39\rlap{.}^\mathrm{s}231$ & $40^{\circ}56'37\rlap{.}''68$ & & & \\
G79.30+0.28 & $ 20^\mathrm{h}32^\mathrm{m}29\rlap{.}^\mathrm{s}479$ & $40^{\circ}16'04\rlap{.}''64$ & & & \\ 
\hline
\end{tabular}
\end{center}
$^{\mathrm{a}}${At 21.96~GHz}
\end{table}
\end{landscape}

\begin{landscape}
\begin{table}[ht]
\footnotesize
\begin{center}
\vspace{-1.7cm}
\caption{Parameters of the VLA maps\label{maps}  }
\begin{tabular}{lcccccc}
\hline
\hline
& \multicolumn{2}{c}{Natural weighted maps} & &  \multicolumn{2}{c}{Uniform weighted maps without short spacings} & \\\cline{2-3}\cline{5-6}
& Beamsize  & map $rms$ noise  & LAS$^{\mathrm{a}}$ & Beamsize& map $rms$ noise &   \\
Source   &  ($'' \times ''$; $^\circ$) & ($\mu$Jy bm$^{-1}$) & ($''$) &  ($'' \times ''$; $^\circ$) & ($\mu$Jy bm$^{-1}$) & $N_\mathrm{sources}^{\mathrm{b}}$\\
\hline
G5.89-0.39 &   $0.190 \times 0.096$; -16.5 & 1700  & -- & -- &-- & 0  \\
G5.97-1.17  &  $0.190 \times 0.093$; -14.6 & 90 & 1.0 & $0.150 \times 0.051$; -10.7 & 47 & 3   \\
G19.61-0.23  &   $0.153 \times 0.094$; -21.9 & 400   & -- & -- &-- & 0 \\
G20.08-0.14 &    $0.153 \times 0.094$; -21.9 & 805   & 0.25 & $0.107 \times 0.061$; -22.1$^{\mathrm{c}}$ & 460$^{\mathrm{c}}$ & 1  \\
G28.29-0.36  &  $0.137 \times 0.093$; -26.2 & 240 & 0.4 & $0.082 \times 0.066$; -27.8& 40 & 1 \\
G35.20-1.74 &  $0.128 \times 0.093$; -31.8 & 419 & 0.4 & $0.078 \times 0.068$; -63.1& 60 & 1\\
G60.88+0.13 & $0.135 \times 0.100$; -64.8 & 47 & 0.4 & $0.093 \times 0.060$; 89.3& 31 & 1\\
G61.48+0.09A &  $0.125 \times 0.091$; -63.8 & 261 & 0.4 & $0.089 \times 0.056$; -89.3& 30 & 2 \\
G76.18+0.13$^{\mathrm{d}}$ &  $0.135 \times 0.094$; -72.4 & 15 & --  & $0.094 \times 0.057$; -85.8 & 28 & 1 \\
G76.38-0.62$^{\mathrm{d}}$ & $0.133 \times 0.093$; -73.7 & 180 & --  & $0.095  \times 0.057$; -85.3& 55 & 1 \\ 
G78.44+2.66$^{\mathrm{d}}$ & $0.134 \times 0.100$; -78.8 & 26   & -- &$0.092 \times 0.058$; 78.3  & 24 & 2 \\
G79.30+0.28$^{\mathrm{d}}$ & $0.138\times 0.099$; -76.0 & 15   & -- & -- &-- & 0 \\ 
\hline
\end{tabular}
\end{center}
$^{\mathrm{a}}${Largest angular scale mapped resulting from removing the short spaced baselines from the $uv$ dataset.} \\ 
$^{\mathrm{b}}${Number of compact sources associated with the region.} \\ 
$^{\mathrm{c}}${For this region, the map without the shortest baselines was obtained with natural weighting.} \\ 
$^{\mathrm{d}}${We only detect compact sources because any extended structure in the constructed map is filtered out even using the total uvrange.} 
\end{table}
\end{landscape}

\begin{landscape}
\begin{table}[ht]
\footnotesize
\begin{center}
\vspace{-1.7cm}
\caption{Parameters of the compact sources \label{sources}$^{\mathrm{a}}$}
\begin{tabular}{p{2.5cm}cccccccc}
\hline
\hline
& \multicolumn{2}{c}{Coordinates}    &   $\theta_M$  x  $\theta_m$  ;  P.A. $^{\mathrm{b}}$ & $S_\mathrm{1.3cm}$  & $I^{Peak}_\mathrm{1.3cm}$ & Adopted\\
Region   &  $\alpha$~(J2000) & $\delta$~(J2000)    &  ($mas  \times  mas$  ;   $^\circ$) &  (mJy)  &  (mJy bm$^{-1}$) & Name  \\
\hline
G5.97-1.17 & $18^\mathrm{h}03^\mathrm{m}40\rlap{.}^\mathrm{s}506$ & $-24^{\circ}22'44\rlap{.}''33$ &  $ 255 \pm 12  \times 126 \pm 12  ;  23 \pm $ 5& $ 8.0 \pm 0.4$  & $ 1.40 \pm 0.06$ & G5.97 proplyd$^{\mathrm{c}}$\\
G5.97-1.17 & $18^\mathrm{h}03^\mathrm{m}40\rlap{.}^\mathrm{s}345$ & $-24^{\circ}22'42\rlap{.}''85$ &    Unresolved or poorly determined  & $ 1.3 \pm 0.1$ & $ 0.94 \pm 0.05$& G5.97-Her 36 SE\\
G5.97-1.17 & $18^\mathrm{h}03^\mathrm{m}40\rlap{.}^\mathrm{s}360$ & $-24^{\circ}22'39\rlap{.}''85$ &      Unresolved or poorly determined      & $ 1.16 \pm 0.11$  & $ 0.97 \pm 0.05$& G5.97-Her 36N\\ 
G20.08-0.14 & $18^\mathrm{h}28^\mathrm{m}10\rlap{.}^\mathrm{s}280$ & $-11^{\circ}28'47\rlap{.}''94$ &  $ 150 \pm 40 \times 70 \pm 20  ;  180 \pm $ 20 & $ 10.0 \pm 1.7$ & $ 3.6 \pm 0.5$& G20.08-VLA1\\
G28.29-0.36 & $18^\mathrm{h}44^\mathrm{m}15\rlap{.}^\mathrm{s}109$ & $-4^{\circ}17'55\rlap{.}''34$ &  $ 90 \pm 10 \times 80 \pm 20  ;  50 \pm $ 60   & $ 2.43 \pm 0.14$  & $ 1.03 \pm 0.07$&G28.29-VLA1\\
G35.20-1.74 & $19^\mathrm{h}01^\mathrm{m}46\rlap{.}^\mathrm{s}460$ & $01^{\circ}13'23\rlap{.}''60$  &   $ 67 \pm 3 \times 39 \pm 3  ;  131 \pm $ 5     & $ 5.24\pm 0.14$ & $ 3.45 \pm 0.06$&G35.20-VLA1\\
G60.88+0.13 & $19^\mathrm{h}46^\mathrm{m}20\rlap{.}^\mathrm{s}016$ & $24^{\circ}35'28\rlap{.}''79$  &   Unresolved or poorly determined      & $ 0.33 \pm 0.05$& $ 0.34 \pm 0.03$ &G60.88-VLA1\\
G61.48+0.09A & $19^\mathrm{h}46^\mathrm{m}49\rlap{.}^\mathrm{s}123$ & $25^{\circ}12'45\rlap{.}''19$ & $ 170 \pm 20 \times 64 \pm 13  ;  162 \pm $ 5 & $ 2.4 \pm 0.2$& $ 0.62 \pm 0.04$&G61.48-VLA1\\
G61.48+0.09A & $19^\mathrm{h}46^\mathrm{m}49\rlap{.}^\mathrm{s}092$ & $25^{\circ}12'45\rlap{.}''00$ &  Unresolved or poorly determined    & $ 0.4 \pm 0.07$ & $ 0.32 \pm 0.04$&G61.48-VLA2\\
G76.18+0.13 & $20^\mathrm{h}23^\mathrm{m}45\rlap{.}^\mathrm{s}744$ & $37^{\circ}38'32\rlap{.}''46$ &  Unresolved or poorly determined   & $ 2.38 \pm 0.06$ & $ 2.21 \pm 0.03$&G76.18-VLA1\\
G76.38-0.62 & $20^\mathrm{h}27^\mathrm{m}26\rlap{.}^\mathrm{s}774$ & $37^{\circ}22'47\rlap{.}''75$ &  $ 96 \pm 2 \times 54 \pm 1 ;  108 \pm $ 2 & $ 21.7 \pm 0.3$ & $ 10.90 \pm 0.09$&G76.38-VLA1\\
G78.44+2.66 & $20^\mathrm{h}19^\mathrm{m}38\rlap{.}^\mathrm{s}849$ & $40^{\circ}56'36\rlap{.}''68$ &   Unresolved or poorly determined  & $ 0.21 \pm 0.04$& $ 0.24 \pm 0.02$&G78.44-VLA4$^{\mathrm{d}}$ \\
G78.44+2.66 & $20^\mathrm{h}19^\mathrm{m}39\rlap{.}^\mathrm{s}600$ & $40^{\circ}56'31\rlap{.}''93$ & Unresolved or poorly determined  & $  0.24 \pm 0.06$   & $  0.14 \pm 0.02$&G78.44-VLA1\\
\hline
\end{tabular}
\end{center}
$^{\mathrm{a}}${Obtained from the uniform weighted maps shown in Figures \ref{compact_a} and \ref{compact_b} whose parameters are listed in Table \ref{maps}.}\\ 
$^{\mathrm{b}}${The angular size is deconvolved from the synthesized beam of the corresponding map (see Table \ref{maps}).} \\ 
$^{\mathrm{c}}${Name taken from \citet{masque2014}.}\\ 
$^{\mathrm{d}}${Name taken from \citet{neria2010}}. 
\end{table}
\end{landscape}

\begin{landscape}
\begin{table}[ht]
\footnotesize
\begin{center}
\vspace{-1.7cm}
\caption{Derived physical parameters of the compact sources \label{physical_parameters}$^{\mathrm{a}}$}
\begin{tabular}{p{3.1cm}ccccccc}
\hline
\hline
              & $T_\mathrm{B}^{\mathrm{b}}$ & & $EM^{\mathrm{d}}$  & $n_\mathrm{e}^{\mathrm{e}}$ & $\theta_\mathrm{eq}^{\mathrm{f}}$ & $log(\dot{N})^{\mathrm{g}}$ & Size\\
Source   &  (K) & $\tau^{\mathrm{c}}$  & (10$^8$ cm$^{-6}$ pc$^{-1}$) & (10$^5$ cm$^{-3}$)  &  ($\arcsec$) & s$^{-1}$ &(AU $ \times $ AU )    \\
\hline
G5.97-Proplyd & $900 \pm 100$& $0.10 \pm 0.01$ &         $1.9 \pm 0.2$     &  $4.1 \pm 0.3$     &  $0.18 \pm 0.01$   &  45.05        &   $ 330 \pm 20  \times 160 \pm 20 $  \\
G5.97-Her 36 SE &$\geq600$ & $\geq0.06$&            $\geq1.2$  	&  $\geq4.8$		&  $\leq0.09$  &  44.23						&$ \leq200 \times \leq70 $\\
G5.97-Her 36N &$\geq600$ & $\geq0.06$&           $\geq1.1$ 	&   $\geq4.6$	&  $\leq0.087$ &   44.20 						&$ \leq 200  \times \leq70 $\\
G20.08-VLA1 &$3300 \pm 1500$ & $ 0.4 \pm 0.2$&     $8\pm 4$  	&  $6.9 \pm 1.3$	&  $0.11 \pm 0.02$ & 46.22         &$ 520 \pm 130  \times 180 \pm 80 $\\
G28.29-VLA1&$1200 \pm 300$& $ 0.13 \pm 0.04$   &       $2.6 \pm 0.8$ 	&  $4.4 \pm 0.7$	&  $0.09 \pm 0.011$  &  45.34       &$ 300 \pm 50  \times 270\pm 60 $\\
G35.20-VLA1&$7400 \pm 700$ & $ 1.3 \pm 0.3$    &     $27\pm 5$	&  $18.4 \pm 1.8$	&  $0.051 \pm 0.002$   &  45.89      &$ 210 \pm 10  \times 130 \pm 10 $\\
G60.88-VLA1 & $\geq200$& $ \geq0.02$ &	$ \geq0.4$		 &  $\geq2.4$ 		&  $\leq0.08$  &  44.15   				&$ \leq200  \times \leq100 $\\
G61.48-VLA1 &$800 \pm 200$ & $ 0.08 \pm 0.02$&	$1.7 \pm 0.4$ &  $4.1\pm 0.6$	&  $0.104 \pm 0.012$ & 44.89   		&$ 340 \pm 40  \times 130 \pm 30 $\\	
G61.48-VLA2 &$\geq2000$ & $ \geq0.3$&	$\geq5.0$  &  $\geq8.5$ &  $\leq0.07$  &  45.38  						&$\leq200 \times  \leq100 $\\			
G78.44-VLA4 &$\geq600$ & $ \geq0.06$&   $\geq1.1$  &  $\geq3.9$	&  $\leq0.07$ & 44.06 &$ \leq300  \times \leq 200 $ \\
G78.44-VLA1&$\geq100$ & $\geq0.02$  &  $\geq0.3$ &  $\geq2.1$	&  $\leq0.07$  & 43.85&$ \leq300  \times  \leq 200 $ \\
\hline
\end{tabular}
\end{center}
$^{\mathrm{a}}${The limits correspond to unresolved sources. For these cases, we adopted the synthesized beam as the source size.} \\
$^{\mathrm{b}}${Brightness temperature obtained with the expression $\left [ \frac{T_\mathrm{B}}{K} \right ] =3.68 \left [ \frac{S_\mathrm{1.3cm}}{\mathrm{mJy}}  \right ] \left [ \frac{\theta_\mathrm{M}}{\mathrm{arcsec}}  \right ]^{-1}  \left [ \frac{\theta_\mathrm{m}}{\mathrm{arcsec}}  \right ]^{-1}$ where $\theta_M$, $\theta_m$ and $S_\mathrm{1.3cm}$ are given in Table \ref{sources}.}\\ 
\vspace{0.3cm}
$^{\mathrm{c}}${Optical depth derived from $T_\mathrm{B}$ assuming an electron temperature of $T_\mathrm{e}=10^4$ K} \\ 
\vspace{0.3cm}
$^{\mathrm{d}}${Emission measure obtained with the equation $\left [ \frac{EM}{\mathrm{cm^{-6} pc}} \right ] =12.195 \left [ \frac{T_\mathrm{e}}{\mathrm{K}} \right ]^{1.35}  \tau   \left [ \frac{\nu}{\mathrm{GHz}} \right ]^{2.1}  $ where we followed the approximation of \citet{altenhoff1960} for radio frecuencies and assumed $T_\mathrm{e}=10^4$ K. } \\
$^{\mathrm{e}}${Volumic density of ionized material, derived assuming fully ionized volume and adopting the source size as the path length of the emitting region which yields to the expression  $ \left [ \frac{n_\mathrm{e}}{\mathrm{cm^{-3}}} \right ] =  454.08 \left [ \frac{EM}{\mathrm{cm^{-6} pc}} \right ]^{0.5} \left [ \frac{\theta_\mathrm{eq}}{\mathrm{arcsec}} \right ]^{-0.5}   \left [ \frac{D}{\mathrm{pc}} \right ]^{-0.5}$. Thus, we assumed cylindrical morphology}\\
$^{\mathrm{f}}${Equivalent source size derived from $\theta_\mathrm{eq}=(\theta_\mathrm{M}\theta_\mathrm{m})^{1/2}$}. The upper limits are derived from the synthesized beam.\\
$^{\mathrm{g}}${Lyman continuum photon flux required to maintain ionization of the source $\left [ \frac{\dot{N}}{\mathrm{s}^{-1}} \right ] = 1.75 \times 10^{39} \left [ \frac{\theta_\mathrm{eq}}{\mathrm{arcsec}} \right ]^3 \left [ \frac{n_\mathrm{e}}{\mathrm{cm^{-3}}} \right ]^2 \alpha \left [ \frac{D}{\mathrm{pc}} \right ]^3$, where $\alpha$ is the recombination coefficient that, at $T_\mathrm{e}=10^4$ K, is $3 \times 10^{-13}$ cm$^{-3}$ s$^{-1}$. This equation implicitly takes into account the optical depth.}\\
\end{table}
\end{landscape}


\end{document}